\documentstyle[12pt]{article}
\begin{document}
\thispagestyle{empty}
\begin{titlepage}
\title{High order behaviour of perturbation recursive relations}

\author{
   O.Yu. Shvedov  \\
{\small{\em Institute for Nuclear Research of the Russian
Academy of Sciences,  }}\\ {\small{\em 60-th October Anniversary Prospect
7a, Moscow 117312, Russia
}}\\ {\small and}\\ {\small{\em Moscow State University }}\\
{\small{\em Vorobievy gory, Moscow 119899, Russia}} }

\end{titlepage}
\maketitle

\begin{center}
{\bf Abstract}
\end{center}

The problem of large order behaviour of perturbation theory for quantum
mechanical systems is considered. A new approach to it is developed.
An explicit mechanism showing the connection between large order recursive
relations and classical euclidean equations of motion is found. Large
order asymptotics of the solution to the recursive relations is constructed.
The developed method is applicable to the excited states, as well as to
the ground state. Singular points of the obtained asymptotics of the
perturbation series for eigenfunctions and density matrices are investigated
and formulas being valid near such points are obtained.

\newpage

\section{Introduction}

A factorial growth of perturbation series coefficients which implies the
divergence of the expansion \cite{D} has been studied in many papers (see,
for example, \cite{L,BW2,BBW,BLGZJ,BPZJ,ZJ,RS}). The corresponding large
order asymptotics is usually constructed by representing the $k$-th order of
perturbation theory as a functional integral and applying the saddle-point
technique to the obtained formula. Therefore, high order behaviour of
peturbation theory is specified by saddle points expressed in terms of
classical euclidean solutions.

Examples of emloyment of this method are:

i) calculating large order asymptotics for quantities like euclidean Green
functions both in quantum mechanics and quantum field theory \cite{L};

ii) finding high order behaviour of energy level perturbation theory in
quantum mechanics \cite{BLGZJ,BPZJ,ZJ,RS};

iii) evaluating the asymptotics of ground state eigenfunction perturbation
theory as the number of order $k$ tends to infinity and the wave function
argument is of order $\sqrt{k}$ \cite{S2}.

It is the form
\begin{equation}\label{1}
{\cal H}(\sqrt{g}\hat{p},\sqrt{g}\hat{q})/g
\end{equation}
of the dependence of the Hamiltonian on the perturbation theory parameter $g$
that allows us to make use of the saddle-point method and to clarify the role
of classical euclidean solutions. Here $\hat{p}$ is a momentum operator,
$\hat{q}$ is a coordinate operator, the behaviour of the Hamiltonian as its
argument tends to zero is like $\hat{p}^2/2+\hat{q}^2/2$ (the harmonic
oscilator Hamiltonian), so there are no terms of order $1/g,1/\sqrt{g}$.
If the form of the Hamiltonian differs from (\ref{1}), one can separate it
into a leading (''classical'') part having the form (\ref{1}) that determines
classical euclidean equations of motion and a correction (''quantum part'')
being of a form like $g^{\alpha}f(\sqrt{g}\hat{p},\sqrt{g}\hat{q}),
\alpha\ge 0,$ which can influence only a pre-exponential factor
(some examples are given in \cite{ZJ,RS}).

However, one can develop the perturbation theory not only by making use of the
path integral representation for corresponding quantities. As a rule, when one
investigates quantum mechanical case and, in particular, checks the large
order asymptotics by numerical calculation \cite{BW1}, one applies another
technique, namely, one studies the perturbation recursive relations expressing
the $k$-th order of perturbation theory for the eigenvalue and the
eigenfunction in terms of preceding orders. Large orders of the recursive
relations have not been analyzed except for the case of the quartic potential,
${\cal H}=\hat{p}^2/2+\hat{q}^2/2-g\hat{q}^4$, see \cite{BW2,BBW}.
The explicit mechanism showing the correspondence between recursive relations
and euclidean classical solutions has not been considered yet. It is studied
in this paper.

The main idea of the method is the following. First, perturbation recursive
relations is written in terms of the operator of differentiation $\partial/
\partial k$ with respect to the perturbation theory order $k$. Namely, the
expression for the $k$-th order $\Psi_k$ of the wave function perturbation
expansion contains the preceding orders $\Psi_{k-p}$ which can be expressed
as $e^{-p\frac{\partial}{\partial k}}\Psi_k.$ Second, consider large order
behaviour of recursive relations, i.e. let $k$ be proportional to some large
parameter $N$. As it is shown in section 2, there exists such rescaling of the
wave function argument (being $n$-dimensional in the case of the
$n$-dimensional quantum mechanics) and of the eigenfunction that after it
the coefficients of the derivation operators become of order 1/N. This
allows us to find the asymptotic solution to the modified equation in a form,
analogous to the tunnel semiclassical approximation, namely, $e^{-NS}$,
where the function $S$ depending on the rescaled coordinates and on the
ratio $k/N$ satisfies the equation of a Hamilton-Jacobi type. As it is known,
for solving such equation, one must consider the corresponding Hamiltonian
system. As the number of arguments of the function $S$ is equal to $n+1$
(one more coordinate, $k/N$, is added), the classical Hamiltonian system
corresponding to recursive relations contains one more coordinate and one
more canonically conjugated momentum. It occurs that the solutions to this
$(n+1)$-dimensional Hamiltonian system can be expressed in terms of the
solutions to an $n$-dimensional Hamiltonian system associated with the
euclidean equation of motion (see section 2 for more details).

In order to find the solution to Hamilton-Jacobi equation, one must know
boundary conditions for the function $S$. For finding such conditions, one
can make use of the following observation. The $k$-th order of the
eigenfunction perturbation
 theory can be presented as a product of a polynomial
function by a Gaussian exponent (it is the ground state wave function of the
non-perturbated system being a harmonic oscilator). The polynomial function
can be approximated at large $x$ by a leading power function. The coefficient
of it is calculable explicitly. On the other hand, an asymptotic formula must
be valid, in particular, at large values of an argument. Section 3 contains a
comparison of these two formulas and obtaining a requirement for the function
$S$, as well as for the pre-exponential factor, as their argument tends to
infinity. This allows us to solve corresponding equation. Section 4 deals
with such calculation.

The obtained formulas being applicable not only to the ground state
perturbation theory but also to the excited state perturbaton theory are
valid when the eigenfunction argument is of order $\sqrt{k}$. When it is
fixed, the asymptotics has not a saddle-point form \cite{S2}, but,
nevertheless, can be obtained by recursive relations analysis. Such derivation
is presented in section 5.

Section 6 contains finding the densiy matrix large order asymptotics and
calculation of some matrix elements. Section 7 deals with concluding remarks.

\section{Perturbation recursive relations and Hamiltonian systems}

The purpose of this section is to show a connection between perturbation
recursive relations associated with the Hamiltonians having the dependence
(\ref{1}) on the perturbation theory parameter and the Hamiltonian system
associated with the classical analog of eq.(\ref{1}).

Consider for the simplicity the one-dimensional quantum mechanical case,
choose the Hamiltonian to be a sum of kinetic and potential terms and
suppose the potential to be even, so
\begin{equation}\label{2}
{\cal H}=-\frac{1}{2}\frac{d^2}{dx^2}+\frac{1}{g}V(\sqrt{g}x)=
-\frac{1}{2}\frac{d^2}{dx^2}+\frac{x^2}{2}+ga_4x^4+g^2a_6x^6+...
\end{equation}
Examine the perturbation theory for the equation ${\cal H}\Psi=E\Psi.$ Both
eigenvalue and eigenfunction can be expanded into asymptotic series,
$$
E_n=\sum g^k E_{n,k}, \Psi_n(x)=\sum g^k \Psi_{n,k}(x).
$$
Here $n$ is a number of energy level. Notice that the wave function is
expanded at fixed value of its argument $x$. For the zeroth order of
perturbation theory we have an oscilator equation,
$$
(-\frac{1}{2}\frac{d^2}{dx^2}+\frac{x^2}{2}-E_{n,0})\Psi_{n,0}=0.
$$
The $n$-th energy level is equal to $E_{n,0}=n+1/2$, while the wave function
has the form of a product of the $n$-th degree Hermite polynomial by the
Gaussian exponent,
$$
\Psi_{n,0}=const H_{n}(x) e^{-x^2/2}.
$$
Let us choose the constant in this formula in order to make the coefficient
of the leading power $x^n$ to be equal to 1, so large $x$ behaviour of the
$n$-th state eigenfunction is as follows,
\begin{equation}\label{3}
\Psi_{n,0}(x)\sim x^ne^{-x^2/2}, x\rightarrow \infty.
\end{equation}
The perturbation theory can be developed by the following recursive relations
obtained from eq.(\ref{2}),
\begin{equation}
\left(-\frac{1}{2}\frac{d^2}{dx^2} + \frac{x^2}{2} \right)
 \Psi_{n,k}(x)
 + a_4 x^4 \Psi_{n,k-1}(x)+...=\sum_{j=0}^{k} E_{n,j} \Psi_{n,k-j}(x).
\label{4}
\end{equation}
Notice that $\Psi_{n,k}$ is defined from this relation up to a factor
$const \Psi_{n,0}$, so one must impose an additional condition on
$\Psi_{n,k}$, for example, the following,
\begin{equation}
\frac{d^n}{dx^n}(e^{x^2/2}\Psi_{n,k})(x=0)=0, k\ge 1.
\label{5}
\end{equation}
Consider the construction of high order asymptotics. Suppose $k$ to be large,
i.e. to be proportional to some large perameter N, $k=N\kappa$. The aim is
to find such rescaling of coordinate $x$ and function $\Psi_{n,k}$ that
provides a possibility of searching the asymptotics in a form $e^{-NS(q)}$,
where $q$ is a set of rescaled coordinates and $\kappa$. Such rescaling
of $x$ is $x=\sqrt{N}\eta$, namely, the expression
$(-d^2/dx^2+x^2)e^{-NS(\kappa,\eta)}$
is then approximately equal to
$N(-(\partial S/\partial \eta)^2+{\eta}^2)e^{-NS(\kappa,\eta)}$, i.e.
these two terms are of the same order. To make other terms in the left-hand
side of eq.(\ref{4}) of this order it is necessary to suppose that
$\Psi_{n,k}$ is approximately in $N$ times greater than $\Psi_{n,k-1},$ so
we come to the following substitution,
\begin{equation}
k=N\kappa, x=\sqrt{N}\eta,
\Psi_{n,k}=N^k\phi_n(\kappa,\eta)\sim N^k e^{-NS(\kappa,\eta)},
N\rightarrow \infty.
\label{6}
\end{equation}
These relations show us that one can find large $k$ asymptotics as
$x/\sqrt{k}\rightarrow const.$ The latter condition coincides with
obtained in \cite{S2} by the path integral technique for the ground state.

We can notice that all the terms of the left-hand side of recursive
relations (\ref{4}) play an important role in constructing large order
asymptotics because of the dependence (\ref{1}) of the Hamiltonian.
If there were terms such as $g^2bx^4$ (for example, if the potential
had a form $x^2/2+g^2(bx^4+ax^6))$ then these terms would not influence a
leading order approximation for $S$. This is in agreement with \cite{ZJ}.

One can also expect that $E_{n,k}$ asymptotics has the form,
\begin{equation}
E_{n,k} \sim N^k e^{-NS(0,\kappa)},
\label{7}
\end{equation}
where the variable $\eta$ is substituted by zero. Namely, function
$\Psi_{n,k}$ is presented as a product of a polynomial function by a Gaussian
exponent,
\begin{equation}
\Psi_{n,k}=\sum A_{n,k,l} x^l e^{-x^2/2}.
\label{8}
\end{equation}
As the condition (\ref{5}) is chosen, $A_{n,k,n}=0.$ It is easy to obtain
from eq.(\ref{4}) that
$E_{n,k}=(n+1)(n+2)A_{n,k,n+2}/2+\sum_{p=4}^{n}a_pA_{n,k,n-p},$
i.e. $E_{n,k}$ is expressed as a finite sum (the number of terms does not
depend on $k$) containing the derivatives of finite order (which also does
not depend on $k$) of the function $\Psi_{n,k}(x)e^{x^2/2}$ as its
argument is equal to zero. As the $\Psi_{n,k}$ asymptotics has the form
(\ref{6}), we can make a limit $\eta \rightarrow 0$ and come to eq.(\ref{7}).

It will be shown that $S(\eta,\kappa) > S(0,\kappa)$. This implies that we
can omit the terms in the right-hand side of eq.(\ref{4}) as $j\sim k$
because they are exponentially small in comparison with the each order of
the left-hand side. This omission is not valid only if $q$ belongs to
a small region near zero which decreases as $k$ increases.

Notice also that if $j\sim 1$ then the corresponding term in the right-hand
side of eq.(\ref{4}) is of order $1/N^{j+1}$ in comparison with all terms
in the left-hand side and, therefore, can influence only  a pre-exponential
factor and corrections. Thus, the function $S$ is determined only by
left-hand side terms and does not depend on a number of energy level.

We can also rewrite quantities such as $\Psi_{n,k-p}$ as
$e^{-p\partial/\partial k}\Psi_{n,k}=e^{-\frac{p}{N}
\frac{\partial}{\partial \kappa} }\Psi_{n,k}$ and, therefore, come to the
following form of recursive relations,
\begin{equation}
[H(q,-\frac{1}{N}\frac{\partial}{\partial q})-\frac{1}{N}E_{n,0}]
\phi_{n}(\eta,\kappa)=0,
\label{9}
\end{equation}
where $q$ is a set of $\eta$ and $\kappa$, while
\begin{equation}
H(q,-\frac{1}{N}\frac{\partial}{\partial q})=
-\frac{1}{2N^2} \frac{\partial^2}{\partial \eta^2} +
e^{\frac{1}{N}\frac{\partial}{\partial \kappa}}
V(\eta e^{-\frac{1}{2N}\frac{\partial}{\partial \kappa}}).
\label{10}
\end{equation}
The terms of order lesser than $1/N$ are omitted in eq.(\ref{9}). The
generalization of formula (\ref{9}) to multidimensional case is
straightforward.

Let us look for the $\phi$ asymptotics in a form \cite{M},
\begin{equation}
\phi_n(q)=N^a\chi_n(q)e^{-NS(q)}(1+O(1/N)),
\label{11}
\end{equation}
where $a$ is a parameter of order $O(1)$. One can obtain the Hamilton-Jacobi
equation for $S$,
\begin{equation}
H(q, \partial S/\partial q)=0.
\label{12}
\end{equation}
In order to solve it, one must consider the corresponding classical
Hamiltonian system. In the present case, the classical Hamiltonian function
depends on two coordinates $\eta$ and $\kappa$, two momenta, $p_{\eta}$
and $p_{\kappa}$, and can be obtained from eq.(\ref{10}) by substituting
the operators $-N^{-1}\partial/\partial \eta$ and
$-N^{-1}\partial/\partial \kappa$ by $p_{\eta}$ and $p_{\kappa}$
 correspondingly,
\begin{equation}
H(\eta,\kappa,p_{\eta},p_{\kappa})=\frac{1}{2} p_{\eta}^2
 - e^{-p_{\kappa}}
V(\eta e^{p_{\kappa}/2}).
\label{13}
\end{equation}
Let us consider it in more details. First of all, the function (\ref{13})
does not depend on the coordinate $\kappa$, so the momentum $p_{\kappa}$
is an integral of motion. Then, the quantities
\begin{equation}
Q=\eta e^{p_{\kappa}/2}, P=p_{\eta} e^{p_{\kappa}/2}
\label{13*}
\end{equation}
satisfy the euclidean equations of motion,
\begin{equation}
\dot{Q}=P,\dot{P}=V^{'}(Q).
\label{14}
\end{equation}
Thus, the examination of the Hamiltonian system corresponding to the
perturbative recursive relations can be reduced to the study of the
Hamiltonian system associated with moving of a particle in the inverted
potential. If the latter Hamiltonian system is integrable
(for example, in one-dimensional case), the Hamiltonian
system (\ref{14}) is integrable, too. The equation for $\kappa$ has the
form,
\begin{equation}
\dot{\kappa} = e^{-p_{\kappa}}\left[1-\frac{Q}{2}\frac{d}{dQ}\right]V(Q)
\label{15}
\end{equation}
 and can be solved exactly.

For finding large order asymptotics of matrix elements, it is useful to know
high order behaviour of perturbation theory for quantities like
$\rho^{m,n}(x,y)=\Psi_m(x)\Psi_n(y).$ It can be found by making use of the
$\Psi_{n,k}$ asymptotics (this technique is developed in section 6).
But another way of constructing $\rho^{m,n}(x,y)$ large order asymptotics
is to apply technique developed in this section. Namely, the function
$\rho^{m,n}(x,y)$ obeys the equation,
\begin{equation}
[-\frac{1}{2}\frac{\partial^2}{\partial x^2}
-\frac{1}{2}\frac{\partial^2}{\partial y^2}
+\frac{1}{g}V(\sqrt{g}x)
+\frac{1}{g}V(\sqrt{g}y)
-E_m-E_n]\rho^{m,n}(x,y)=0.
\label{16}
\end{equation}
One can develop the perturbation theory
$\rho^{m,n}(x,y)=\sum g^k \rho^{m,n}_{k} (x,y)$, change the variables
$k=N\kappa, x=\sqrt{N}\eta_1, y=\sqrt{N}\eta_2, \rho_{k}^{m,n}=N^k R_{m,n},$
rewrite the recursive relations corresponding to eq. (\ref{16}) in a
way analogous to eq.(\ref{9}), obtain a Hamilton-Jacobi equation and
reduce it to a Hamiltonian system that contains three coordinates
$x,\eta_1,\eta_2$, three momenta $p_{\kappa},p_{\eta_1},p_{\eta_2}$.
The corresponding Hamiltonian function is then
$$
H=p_{\eta_1}^2/2+p_{\eta_2}^2/2
- e^{-p_{\kappa}}V(\eta_1e^{p_{\kappa}/2})
- e^{-p_{\kappa}}V(\eta_2e^{p_{\kappa}/2}).
$$
The corresponding Hamiltonian equations can be solved as well as in previous
case.

One can also examine the perturbation theory for eigenfunctions
in other representations. It occurs that perturbative recursive relations
in any presentation are connected with the Hamiltonian system related with
the investigated one by a canonical transformation. Thus, consideration
of the high orders is associated with the investigation of the euclidean
equations of motion (\ref{14}).

Let us illustrate this declaration by an example.
Consider for the simplicity case of the ground state, i.e. $n=0$.
The function $\Psi_{0,k}$ has then the form (\ref{8}),
$$
\Psi_{0,k}(x)=\sum_{m=1}^{2k} A_{k,m} x^{2m} e^{-x^2/2},
$$
where the polynomial coefficients satisfy the relations:
$$
-(m+1)(2m+1)A_{k,m+1}+2mA_{k,m}+a_4A_{k-1,m-2}+a_6A_{k-2,m-3}+...=
\sum_{j=1}^{k} E_{0,j}A_{k-j,m}.
$$
Assume that $k$ and $m$ are proportional to a large parameter,$k=N\kappa,
m=N\mu,$ and substract the factor $N^{k-m}$ from the coefficients,
$A_{k,m}=N^{k-m}{\cal A}_{k,m}$.
For the reduced coefficients ${\cal A}_{k,m}$ one can
obtain in a leading order a pseudodifferential equation,
$$
H(\mu,\kappa,-\frac{1}{N}\frac{\partial}{\partial \mu},
-\frac{1}{N}\frac{\partial}{\partial \kappa}){\cal A}=0,
$$
where the corresponding Hamiltonian have the form
$$
H(\mu,\kappa,p_{\mu},p_{\kappa})=2(\mu e^{-p_{\mu}/2} - e^{p_{\mu}/2}/2)^2-
e^{-p_{\kappa}}V(e^{(p_{\kappa}+p_{\mu})/2})
$$
and can be obtained from the investigated one by the canonical transformation:
$$
p_{\mu}=2\ln\eta,\mu=\eta(\eta-p_{\eta})/2.
$$
Thus, one can apply the same technique in order to obtain high order
asymptotics in other representations.

\section{Large $x$ behaviour of the $\Psi_{n,k}$ asymptotics and
boundary conditions}

As it has been shown in previous section, the problem of finding the large
 order  asymptotics for the wave function perturbation theory is reduced
 to the problem of solving the Hamilton-Jacobi equation (\ref{12}) and
 the equation for $\chi_n$ which is obtainable from eq.(\ref{9}).
 To solve them, one must know boundary conditions which are to be found
 from other reasoning.

For this purpose, consider the behaviour of $\Psi_{n,k}$ at large $x$.
This allows us to impose boundary conditions on the functions $\chi_n,S$
at large $\eta$.

As $\Psi_{n,k}$ has the form (\ref{8}), large $x$ behaviour of it can be
obtained by approximating the polynomial function by the leading order,
\begin{equation}
\Psi_{n,k}(x) \sim c_{n,k} x^{4k+n} e^{-x^2/2}.
\label{17}
\end{equation}
Quantities $c_{n,k}$ obey recursive relations,$4kc_{n,k}+a_4 c_{n,k-1}=0$,so
$$c_k=(-a_4/4)^k/k!.$$
Let us make use of eq.(\ref{17}) in order to find the conditions to be
imposed on the functions $S,\chi_n$ and the constant $a$. Changing the
variables (\ref{6}) and employing the asymptotic formula for $k!$ at large
$k$,$k!\sim (k/e)^k\sqrt{2\pi k}$, we find that
$$
\phi_n \sim N^{\frac{n-1}{2}}\frac{\eta^n}{\sqrt{2\pi\eta}}
e^{-N\eta^2/2}e^{N\kappa [\ln(-\frac{a_4\eta^4}{4\kappa})+1]}.
$$
Thus, one can expect that
\begin{equation}
a=(n-1)/2,\chi_n \sim \frac{\eta^n}{\sqrt{2\pi\kappa}},
S\sim \frac{\eta^2}{2} - \kappa\ln(-\frac{a_4\eta^4}{4\kappa})-\kappa
\label{18}
\end{equation}
at larges $\eta$. Let us discuss now when the approximation (\ref{18}) works.
In order to answer this question notice that the large parameter $N$ can
be chosen in different ways, it can be rescaled, $N\rightarrow\alpha N$.
Then other variables also rescale, $\kappa\rightarrow\kappa/\alpha$,
$\eta\rightarrow\eta/\sqrt{\alpha}$, $\phi\rightarrow\phi\alpha^{-k}$.
As the function $\Psi_{n,k}$ does not alter, we have
$$
N^{k+a}e^{-NS(\kappa,\eta)}\chi_n(\kappa,\eta)=
(N\alpha)^{k+a}e^{-N\alpha S(\kappa/\alpha,\eta/\sqrt{\alpha})}
\chi_n(\kappa/\alpha,\eta/\sqrt{\alpha}).
$$
Therefore, the following properties of $\chi_n$,$S$
$$
\chi_n(\kappa,\eta)=\alpha^a\chi_n(\kappa/\alpha,\eta/\sqrt{\alpha}),
S(\kappa,\eta)=-\kappa\ln\alpha + \alpha S(\kappa/\alpha,\eta/\sqrt{\alpha}),
$$
which are obviously satisfied for eqs.(\ref{18}), take place. We see that
the values of these functions at fixed $\eta/\sqrt{\kappa}$ are related
with each other. Thus, we notice that if the approximation (\ref{18}) is
valid at some value of $\eta,\kappa$ then it is applicable at all values of
$\eta,\kappa$ having the same ratio $\eta/\sqrt{\kappa}$. Therefore,
$\eta/\sqrt{\kappa}$ is the only parameter determining whether approximation
(\ref{18}) is valid or not. Thus, eq.(\ref{18}) is applicable at
\begin{equation}
\label{19}
\eta/\sqrt{\kappa}\gg 1.
\end{equation}
Taking into account eq.(\ref{18}) as a boundary condition for $S,\chi_n$,
find the solution to the Hamilton-Jacobi equation (\ref{12}) and the
equation for $\chi_n$ obtained also from eq.(\ref{9}).

\section{Large order asymptotics for the eigenfunction perturbation theory}

In this section the functions $S,\chi_n$ and the constant $a$ are found,so
that large order behaviour of the eigenfunction perturbation theory is
obtained.

\subsection{Finding a solution to Hamilton-Jacobi equation}

In order to obtain a solution to the stationary Hamilton-Jacobi equation
let us consider first a non-stationary equation for the function $S(q,t)$
depending on $l$ coordinates $q$ and time parameter $t$,
$$
\frac{\partial S}{\partial t}+H(q,\frac{\partial S}{\partial q})=0,
$$
and remind a helpful property of its solution (see, for example, \cite{M}).
Consider the $2l$-dimensional phase space with first $l$ coordinates being
momenta $p$ and the rest being coordinates $q$. Examine the
$l$-dimensional surface in it,
\begin{equation}
\label{20}
p_i(q)=\frac{\partial S}{\partial q_i}
\end{equation}
depending on $t$ (the arguments $q$,$t$ of the function $S$ are omitted in eq.
(\ref{20})). Consider any point $(p^0,q^0)$ on the surface (\ref{20}) when
$t=0$ and its time evolution, i.e. the trajectory $(p^t,q^t)$ in the
phase space which satisfies the Hamiltonian system,
$$
\dot{p}^{t}_{i}=-\frac{\partial H}{\partial q_i},
\dot{q}^{t}_{i}=\frac{\partial H}{\partial p_i}.
$$
It can be shown \cite{M} that the point $(p^t,q^t)$ belongs to the surface
(\ref{20}) at time moment $t$.

Making use of this property, let us find a solution to the stationary
equation (\ref{12}) which obeys eq.(\ref{18}) under condition (\ref{19}).
This means that the surface (\ref{20}) transforms under this condition to
the surface
\begin{equation}
\label{21}
p_{\kappa}\sim -\ln(-\frac{a_4\eta^4}{4\kappa}),
p_{\eta} \sim \eta(1-\frac{4\kappa}{\eta^2})\sim \eta.
\end{equation}
Reconstruct now the whole surface (\ref{20}), making use of the asymptotics
(\ref{21}). We know that the surface (\ref{20}) is invariant under time
evolution, so we can consider the surface consisting of the trajectories
that satisfy Hamiltonian systam and start from the asymptotical surface
(\ref{21}). Consider any point on it and treat it as an initial condition
for the Hamiltonian system associated with eq.(\ref{13}), which reduces
after substitution (\ref{13*}) to eq.(\ref{14}). An initial condition for
it is as follows,
$$
P\sim Q \sim \sqrt{-\frac{4\kappa}{a_4\eta^4}}
$$
We see that if the condition (\ref{19}) is satisfied then this point $(P,Q)$
is closed to the minimum of the Hamiltonian (\ref{1}),$P=Q=0$, so that the
trajectory should start from it. To countrbalance it, the parameter
$p_{\kappa}$ can take on any value under condition (\ref{19}). Therefore,
the surface (\ref{20}) consists of the following treajectories,
$$
 \eta(p_{\kappa},\tau)=Q(\tau)e^{-p_{\kappa}/2},
 p_{\eta}(p_{\kappa},\tau)=P(\tau)e^{-p_{\kappa}/2},
$$
\begin{equation}
\label{22}
\end{equation}
$$
\kappa(p_{\kappa},\tau)=\lambda (\tau) e^{-p_{\kappa}},
p_{\kappa}(p_{\kappa},\tau)=p_{\kappa},
$$
where $P(\tau),Q(\tau)$ is a solution to eq.(\ref{14}) which starts from
zero as $\tau\rightarrow -\infty$, the parameter $\lambda(\tau)$ is
found from eq.(\ref{15}) and condition $\lambda(-\infty)=0$ derived
from eq.(\ref{21}). It has the form
$$
\lambda(\tau)=\int_{-\infty}^{\tau} d\tau (1-\frac{Q}{2}\frac{d}{dQ})
V(Q(\tau)).
$$
The parameter $\tau$ is defined up to shifting by a constant. In order to
fix it choose the parameter $c$ in the $\tau\rightarrow-\infty$ asymptotics,
$Q(\tau)\sim ce^{\tau}$, to be equal to 1.

Consider an example. Examine the function $V$ in eq.(\ref{2}) which is
depicted in fig.1a. It has a relative minimum at $Q=0$, is positive as
$|Q|<Q_+$ and negative as $|Q|>Q_+$. The euclidean classical solution to eq.
(\ref{14}) starting from $P=Q=0$ as $\tau\rightarrow-\infty$ behaves as
follows (fig.1b). At first stage the coordinate $Q$ increases, then the
 solution reflects from the turning point $Q_+$ and comes back as
 $\tau\rightarrow +\infty$ to point $P=Q=0$. The parameter $\lambda$ has
 then the form
\begin{equation}
\lambda(\tau)=s(\tau)-Q(\tau)\dot{Q}(\tau)/2,
\label{22*}
\end{equation}
where $s(\tau)$ is the action of the euclidean solution,
$s(\tau)=\int_{-\infty}^{\tau} (\dot{Q}^2/2+V(Q)).$ The quantity $\lambda$
has the meaning of the area of the section-lined region in fig.1b. The
corresponding curves at $\kappa-\eta$ space at various values of $p_{\kappa}$
are shown in fig.2. In order to convert from quantities $\kappa,\eta$ to
quantities $p_{\kappa},\tau$ one should first find the parameter $\tau$ from
the relation between it and $\xi=\eta/\sqrt{\kappa}$,
\begin{equation}
\label{23}
\xi=Q(\tau)/\sqrt{\lambda(\tau)}
\end{equation}
and then obtain $p_{\kappa}=\ln(\lambda/\kappa)$. Notice that large values of
$\xi$ correspond to $\tau\rightarrow -\infty$, while small values of $\xi$
are associated with $\tau\rightarrow +\infty$. For the simplicity, consider
the case, when the dependence between $\xi$ and $\tau$ is one-to-one, i.e.
$\dot{\xi}<0$ everywhere. In general case, the solution to Hamilton-Jacobi
equation is multivalued, and it is necessary to apply the Maslov canonical
operator technique \cite{M} for finding the asymptotic solution to
eq.(\ref{9}).

Instead of the parameter $\tau$, one can use the variable $Q$ such as
$Q=Q(\tau)$. The correspondence between $\xi$ and $Q$ being multivalued is
shown in fig.1c.

In order to find the function $S$ satisfying eq.(\ref{20}) consider the
integral
\begin{equation}
\int p_{\kappa}d\kappa +p_{\eta} d\eta
\label{24}
\end{equation}
which does not depend on the curve connecting two points and equals to
\begin{equation}
(s(\tau)+\lambda(\tau)p_{\kappa})e^{-p_{\kappa}} + const
\label{25}
\end{equation}
The derivatives of it with respect to $\kappa$ and $\eta$ are equal to the
functions $p_{\kappa}$ and $p_{\eta}$ after expression of their arguments
$p_{\kappa}$ and $\tau$ through the quantities $\kappa$ and $\eta$.
As the surface (\ref{22}) lies on the surface $H=0$, the function (\ref{25})
obeys the stationary Hamilton-Jacobi equation (\ref{12}). Because of the
boundary conditions (\ref{18}), the constant in eq.(\ref{25}) is equal to
zero. Converting the parameter $p_{\kappa}$ to $\kappa,\tau$, we obtain
that
\begin{equation}
S(\kappa,\eta)=\kappa \left(\frac{s(\tau)}{\lambda(\tau)}+
\ln\frac{\lambda(\tau)}{\kappa(\tau)} \right),
\label{26}
\end{equation}
the argument $\eta$ is to be expressed through the parameter $\tau$ according
to eq.(\ref{23}). Substituting eq.(\ref{26}) to eq.(\ref{11}), making use of
the asymptotic formula for $k!$ at larges $k$ and applying the expression
$x/\sqrt{k}=\xi$, we obtain that
\begin{equation}
\Psi_{n,k}(\xi\sqrt{k}) \sim k!e^{-kA(\xi)},
\label{27}
\end{equation}
where the pre-exponential factor is omitted,
\begin{equation}
A(\xi)=\frac{s(\tau)}{\lambda(\tau)} + \ln \lambda(\tau) -1
\label{28}
\end{equation}
Remember now that the right-hand side of the recursive relations (\ref{4}) has
been neglected. This procedure is valid if $e^{-kA(\xi)}\gg
e^{-kA(0)}e^{-k\xi^2/2}$.
Consider the function $f(\xi)=A(0)+\xi^2/2-A(\xi)$. At $\xi=0$ we have
$f(\xi)=0$, while $f^{'}(\xi)=(Q-\dot{Q})/\sqrt{\lambda}.$ When
$\dot{Q}<0$, this quantity is obviously positive. As we have supposed
$\dot{\xi}$ to be negative, the quantity $\dot{\lambda}$ should be
positive at $\dot{Q}>0$, i.e.  $\dot{Q}^2-Q\ddot{Q}>0$. This means
that $\dot{Q}<Q.$ Thus, the derivative $f^{'}(\xi)$ is positive
everywhere, so that $f(\xi)>0$.  The right-hand side of eq.(\ref{4})
can be omitted when $kf(\xi)\gg 1$,i.e.  \begin{equation} k\xi^2 \gg 1
\label{29} \end{equation}
because at small $\xi$ the function $f$ is of order $\xi^2$.

Exponential asymptotics (\ref{27}) can be also checked by numerical
calculations. Fig.3 contains the graphs of functions
\begin{equation}\label{A}
A_k(\xi)=-\frac{1}{k}\ln |\frac{\Psi_{0,k}(\xi\sqrt{k})}{k!}|
\end{equation}
as $k=10,20,30$ (dashed lines) and function (\ref{28}) (solid line).
The potential has the form
\begin{equation}
V(Q)=Q^2/2-Q^4.
\label{30}
\end{equation}
We see that functions $A_k(\xi)$ indeed tend to $A(\xi)$ at larges $k$.

\subsection{Calculation of the pre-exponential factor}
For finding the pre-exponential factor in eq.(\ref{11}) consider its
substitution to eq.(\ref{9}). The following
equation for $\chi_n$ can be obtained:
\begin{equation}
\left[H(q,\frac{\partial S}{\partial q}
-\frac{1}{N}\frac{\partial}{\partial q})+\frac{1}{N}E_{n,0}\right]\chi=0,
\label{31}
\end{equation}
Expand the function  $H$ into a series in $1/N$ \cite{M}.
 The term of order $O(1)$ vanishes because of the Hamilton-Jacobi equation,
 the term of order $O(1/N)$ determines the equation for function $\chi_n$.
 This equation has the form:
\begin{equation}
\left(-\frac{1}{N}\frac{\partial H}{\partial p_m}
\frac{\partial}{\partial q_m}-
\frac{1}{2N}\frac{\partial^2 H}{\partial p_m \partial p_n}
\frac{\partial^2}{\partial q_m \partial q_n}+E_{n,0}\right)\chi(q)=0,
\label{32}
\end{equation}
where the summation over repeated indices $m,n$ is assumed, the coordinates
$\kappa,\eta$ are denoted by $q_1,q_2$, the momenta $p_{\kappa},p_{\eta}$
are denoted by $p_1,p_2$. Formula (\ref{32}) can be derived by the expansion
of the function $H$ into a series in powers of the momenta and by checking
this
formula for power functions. Let us substitute variables $q$ for parameters
$\tau,p_{\kappa}$ according to (\ref{22}). The operator
$\frac{\partial H}{\partial p_m} \frac{\partial}{\partial q_m}$
transforms then to the operator of differentiation with respect to
$\tau$.

In order to simplify eq.(\ref{28}) notice that the Jacobian of the
substitution $\tau,p_{\kappa}$ by $q$,
\begin{equation}
J(\tau,p_{\kappa})=\frac{D(\eta,\kappa)}{D(\tau,p_{\kappa})}
\label{32*}
\end{equation}
satisfies the equation \cite{M}
$$
\frac{\partial J}{\partial \tau} =
(\frac{\partial^2 H}{\partial p_m \partial p_n}
\frac{\partial^2 S}{\partial q_m \partial q_n}  +
\frac{\partial^2 H}{\partial p_m \partial q_m})J.
$$
Making use of the nullification of the Hamiltonian second derivatives
with respect to
$p_m,q_m$ and substracting the factor  $J^{-1/2}$ from the function $\chi_n$:
$
\chi_n=f_n/\sqrt{J},
$
one finds the equation for the function $f_n$
$$
\frac{\partial f_n}{\partial \tau} = (n+1/2) f_n.
$$
Therefore, the function $f_n$, as well as $\chi_n$, is defined up
to an arbitrary
factor which depends on $p_{\kappa}$ and does not depend on $\tau$.
Let us use the fact that the Jacobian $J$ has the following form in
this case,
$$
J=\exp\left(-\frac{3p_{\kappa}}{2}\right)
\left[-\lambda \dot{Q}+\frac{Q}{2}\dot{\lambda}\right],
$$
and substract the square root of $e^{-3p_{\kappa}/2}$ into a total
$\tau$-independent factor. The function $\chi_n$ has then the form:
\begin{equation}
\chi_n = \frac{e^{\tau (n+1/2)}g(e^{p_{\kappa}})}
{\sqrt{Q\dot{\lambda}/2-\lambda\dot{Q}}},
\label{33}
\end{equation}
To find the function $g$, let us make use of the condition (18) being valid
at large $\xi$, i.e. at $\tau\rightarrow -\infty$. Under these circumstamces
we have:
$$
Q\sim e^{\tau}, V-QV^{'}/2 \sim -a_4Q^4,\lambda \sim -a_4 e^{4\tau}/4,
$$
so that
$$
\chi_n \sim \frac{g(e^{p_{\kappa}})}{\sqrt{-a_4/4}} e^{\tau (n-2)}.
$$
Consider now eq.(\ref{18}) for $\chi_n$ and substitute the variables
$\eta,\kappa$ by $\tau,p_{\kappa}$ according to eq.(\ref{22}). The following
expression can be found,
$$
\chi_n \sim \frac{e^{(n-2)\tau}e^{-\frac{n-1}{2}p_{\kappa}}}
{\sqrt{-a_4/4}\sqrt{2\pi}}.
$$
Comparing two obtained expressions for $\chi_n$, we find that
\begin{equation}\label{34}
g(e^{p_\kappa})=e^{-\frac{n-1}{2}p_{\kappa}}/\sqrt{2\pi}.
\end{equation}
Making use of eq.(\ref{33}) for $\chi_n$, eq.(\ref{34}) for the function $g$,
the relation (\ref{18}) for the constant $a$ and eqs.(\ref{22}), collecting
all factors in the asymptotics (\ref{11}) and applying the asymptotic formula
for $k!$, one can obtain that
\begin{equation}
\Psi_{n,k}(\xi\sqrt{k}) = \frac{e^{(n+1/2)\tau}}
{\sqrt{Q\dot{\lambda}/2-\lambda\dot{Q}}}\frac{k!}{2\pi\sqrt{k}}
(k/\lambda)^{(n-1)/2}e^{-kA(\xi)}(1+O(1/k)).
\label{35}
\end{equation}
 An interesting feature of eq.(\ref{35}) is the absence of
the singularity in the turning point: although
the velocity $\dot{Q}$ vanishes in this point, the Jacobian is finite
because $\dot{\lambda}\ne 0$.

Formula (\ref{35}) can be also checked  numerically. The
graphs of the functions
 \begin{equation}\label{MK}
 M_k(\xi)= \frac{\Psi_{0,k}(\xi\sqrt{k})}{(k-1)!e^{-kA(\xi)}}
 \end{equation}
for the ground state wave function and the potential (\ref{30}), as well as
the function
 \begin{equation}\label{M}
 M(\xi)=\frac{e^{\tau/2}\sqrt{\lambda}}
 {2\pi \sqrt{Q\dot{\lambda}/2-\lambda\dot{Q}}}.
 \end{equation}
are shown in fig.4 at $k=10,20,30,40$.
  Fig.4 shows us that the asymptotics (\ref{35}) is
 good at all values of $\xi$ except some region near the point $\xi=0$.
 The size of this region decreases as $k$ increases.
This is in agreement with eq.(\ref{29}) determining the range of validity
of the obtained asymptotics.

\section{Large order behaviour of $\Psi_{n,k}(x)$ at fixed $x$}

An interesting feature of the asymptotics (\ref{35}) constructed in previous
section is the singularity of the pre-exponential factor near the point
$\xi=0$.
Namely, small values of $\xi$ correspond to $\tau\rightarrow +\infty$.
Find the behaviour of $Q,\lambda,A$ and the pre-exponential factor
in eq.(\ref{35}) at these values of $\tau$. First, notice that the coordinate
$Q$ behaves as $\tau \rightarrow +\infty$ as follows, $Q\sim ce^{-\tau}$. The
value of the constant $c$ can be found by the following procedure.
Consider a small quantity $q_0$.
The classical trajectory $Q(\tau)$ (fig.1b)
passes through this point, $q_0$, at two euclidean
time moments being approximately equal to $\ln q_0$ and $\ln c/q_0$
 because of the $Q(\tau)$ asymptotics
as $\tau\rightarrow\pm\infty$. On the other hand, the difference between
these time moments $\ln c - 2\ln q_0$ is equal to
$2\int_{q_0}^{Q_+} dq/\sqrt{2V(Q)}$, so that
\begin{equation}\label{35c}
c=\exp 2 (\ln Q_+ + \int_0^{Q_+} dq(\frac{1}{\sqrt{2V(q)}}-\frac{1}{q})).
\end{equation}
Making use of eqs.(\ref{22*}),(\ref{23}),(\ref{28}), we obtain
$$
\lambda\sim s(+\infty)-c^2e^{-2\tau},
\xi\sim\frac{ce^{-\tau}}{\sqrt{s(+\infty)}},
A\sim \ln s(+\infty)-\xi^2/2.
$$
Taking into account all factors, find that at small $\xi$ the asymptotics
(\ref{35}) transforms to the following formula,
\begin{equation}
\frac{e^{k\xi^2/2}}{(\xi\sqrt{k})^{n+1}}\frac{c^{n+1/2}}{2\pi}
\frac{k!k^{n-1/2}}{(s(+\infty))^{k+n+1/2}},
\label{36}
\end{equation}
i.e. the pre-exponential factor diverges as $1/\xi^{n+1}$, where $n$ is a
number of level.

On the other hand, each order of the eigenfunction perturbation theory is
non-singular everywhere and, in particular, at the point $\xi=0$.
The difficulty can be resolved as follows. As it has been shown in the
previous section, the obtained asymptotic formula (\ref{35}) may be not valid
at $k\xi^2\sim 1$, i.e. at fixed argument of $\Psi_{n,k}$ because two sides
of eq.(\ref{4}) become of the same order. Therefore, the asymptotics under
conditions
\begin{equation}
k\rightarrow\infty,x=const,
\label{37}
\end{equation}
has another form than eq.(\ref{36}) and is non-singular. It is investigated
in this section and allows us to find the large order asymptotics for
eigenvalues which is also studied in this section.

Let us look for the asymptotics under conditions (\ref{37}) in a form,
\begin{equation}
\Psi_{n,k}(x) \sim \frac{c^{n+1/2}}{2\pi} \frac{k!k^{n-1/2}}
{(s(+\infty))^{k+n+1/2}}X_n(x).
\label{38}
\end{equation}
Take into account that at  $k\xi^2\gg 1,k\xi^3\ll 1$
both eqs.(\ref{36}) and (\ref{38}) are applicable.
This means that the function $X_n$ behaves at large $x$ as follows,
\begin{equation}
X_n(x) \sim e^{x^2/2}/x^{n+1},x\rightarrow\infty.
\label{39}
\end{equation}
Let us also look for the eigenvalue high order asymptotics in the form:
\begin{equation}
E_{n,k}\sim\frac{c^{n+1/2}}{2\pi} \frac{k!k^{n-1/2}}
{(s(+\infty))^{k+n+1/2}}{\cal E}_n
\label{40}
\end{equation}
and consider the substitution of eqs.(\ref{38}),(\ref{40}) to the recursive
relations (\ref{4}). Notice that at fixed $x$ and larges $k$ the first
term at the left-hand side of eq.(\ref{4}) prevails others, while
right-hand
side contains two terms of the same order, $E_{n,0}\Psi_{n,k}$ and
$E_{n,k}\Psi_{n,0}$. Therefore, the following equation is obtained for $X_n$:
\begin{equation}
\left(-\frac{1}{2}\frac{d^2}{dx^2}+\frac{x^2}{2}-n-1/2\right)X_n(x)=
{\cal E}_n\Psi_{n,0}(x),
\label{41}
\end{equation}
It has the form of a harmonic oscilator equation with the right-hand
side.

First  of all, solve eq.(\ref{41}) with the boundary condition (\ref{39})
at $n=0$, so that $\Psi_{0,0}(x)=e^{-x^2/2}$.
Let us search for the solution in a
form:
$$
X_0(x)=Y(x)e^{-x^2/2},
$$
for $Y^{'}(x)$ one can obtain the boundary condition $Y^{'}(x)\sim2e^{x^2}$
and the equation
$$
-(Y^{'}e^{-x^2})^{'}=2e^{-x^2}.
$$
Thus,the solution $X_0$ is the following:
$$
X_0(x)=-\frac{4}{\sqrt{\pi}}e^{-x^2/2} \int_{0}^{x} dx^{'} e^{x^{'2}}
\int_{0}^{x^{'}} dx^{''} e^{-x^{''2}},
$$
the boundary condition (\ref{5}) is taken into account. The quantity
${\cal E}_0$ is equal to $-2/\sqrt{\pi}$.

Functions $X_n$ can be derived from the function $X_0$ with the help
of the creation operator $x-d/dx$. Namely, if the function $X_n$
satisfies eqs.(\ref{41}),(\ref{39}) then the function
$$
X_{n+1}=\frac{1}{n+1}(x-d/dx)X_n
$$
satisfies these relations, too. The constant ${\cal E}_{n+1}$ is then
${\cal E}_{n+1}=2{\cal E}_n/(n+1)$, because the normalizing factor in
$\Psi_{n,0}$
is chosen in order to satisfy the relation (\ref{3}).
Notice also that we can add the factor $C_n\Psi_{n,0}$ to the function $X_n$.
The constant $C_n$ can be fixed by the condition (\ref{5}).
 Thus, we obtain
\begin{equation}
X_n=\frac{1}{n!}(x-d/dx)^nX_0+ C_n\Psi_{n,0},
{\cal E}_n=-\frac{2^{n+1}}{n!\sqrt{\pi}}.
\label{42}
\end{equation}
where
$$
C_n=-\frac{1}{n!}\frac{d^n}{dx^n}
\left[\frac{1}{n!}(x-\frac{d}{dx})^nX_0e^{x^2/2}\right],
$$
the derivative is taken at $x=0$.

The asymptotics for eigenvalues (\ref{38}) coincides with the obtained one
in \cite{BLGZJ,ZJ}.

The asymptotic formula (\ref{36}) can be checked by numerical calculations,
too. Consider the functions
\begin{equation}\label{X}
X_{0,k}(x)=\sum_{l=1}^{2k} \frac{A_{k,l}x^{2l}}{A_{k,1}} e^{-x^2/2},
\end{equation}
which are proportional to $\Psi_{0,k}(x)$ and satisfy the condition
$$
X_{0,k}(x)e^{x^2/2} \sim x^2+..., x\rightarrow 0.
$$
Eq.(\ref{38}) implies that this function tends as $k\rightarrow\infty$
 to the solution to eq.(\ref{39}) as $n=0$ which
 is shown by dashed line in fig.5. Functions $X_{0,k}$ as
$k=10,20,30,40$ are shown by solid lines for the potential (\ref{30}) in
fig.5. The figure shows that these functions indeed tend to the solution
to eq.(\ref{41}).

\section{Large order behaviour of perturbation theory for density matrix
and some matrix elements}

In previous sections large order behaviour of $\Psi_{n,k}$ has been found.
But for calculating high order asymptotics of matrix elements it is
necessary to know such asymptotics for the quantities
\begin{equation}\label{43}
\rho^{n_1n_2}(x_1,x_2)=\Psi_{n_1}(x_1)\Psi_{n_2}(x_2)
\end{equation}
Namely, matrix elements are expressed as a trace of a product of this matrix
by the operator corresponding to the observable. When $n_1=n_2$, the matrix
(\ref{43}) is a density matrix. One can expand eq.(\ref{43}) into a series in
$g$,
$$
\rho^{n_1n_2}(x_1,x_2)=\sum_{k=0}^{\infty} g^k \rho^{n_1n_2}_{k}(x_1,x_2)
$$
and express the coefficients through the quantities $\Psi_{n,k}$,
\begin{equation}\label{44}
\rho_k^{n_1n_2}(x_1,x_2)=\sum_{m=0}^{k}
\Psi_{n_1,m}(x_1)\Psi_{n_1,k-m}(x_2).
\end{equation}
This sum is estimated in subsection 6.1, while the perturbation series
asymptotics for some matrix elements is looked for in subsection 6.2.

\subsection{Calculation of the $\rho_k^{n_1n_2}$ asymptotics}

Let us calculate the sum (\ref{44}) at large $k$. Since the $\Psi_{n,m}(x)$
asymptotics is already known as $m$ is proportional to the large parameter
$N$ and $x\sim\sqrt{N}$, one can estimate the sum (\ref{44}) under conditions
\begin{equation}\label{44*}
k=N\kappa, x_1=\sqrt{N}\eta_1, x_2=\sqrt{N}\eta_2.
\end{equation}
Denoting $m=N\mu$ and making use of eq.(\ref{11}), we obtain:
\begin{equation}\label{45}
\rho^{n_1n_2}_{k}\sim\sum_{N\mu=0}^{N\kappa} N^{k+\frac{n_1+n_2}{2}-1}
\chi_{n_1}(\mu,\eta_1)\chi_{n_2}(\kappa-\mu,\eta_2)
e^{-N(S(\mu,\eta_1)+S(\kappa-\mu,\eta_2))}
\end{equation}
Consider the point $\mu_0$ being the point of minimum of the function
\begin{equation}\label{45*}
B(\mu)= S(\mu,\eta_1)+S(\kappa-\mu,\eta_2)
\end{equation}
and suppose
$\mu_0$ to belong the interval $(0,\kappa)$. The main contribution
to the sum (\ref{45}) is then given by any small interval
$(\mu_0-b/N^{\beta}, \mu_0+b/N^{\beta})$, where $0<\beta<1/2$.
 Therefore, the pre-exponential factors can be replaced by their
values at point $\mu_0$, while the function $B$ can be substituted by
the expression $$
B(\mu)\sim B(\mu_0)+\frac{1}{2}B^{''}(\mu_0)(\mu-\mu_0)^2.
$$
As the distance between neighbour values of $\mu$ is equal to $1/N$, the
sum over $\mu$ can be replaced by the integral $N\int d\mu$. Making use of
these observations and calculating the Gaussian integral over $\mu$,
we obtain that
\begin{equation}\label{46}
\rho^{n_1n_2}_{k}\sim N^{k+\frac{n_1+n_2-1}{2}}
\exp(-NB_0) \sqrt{2\pi/B^{''}(\mu_0)}
\chi_{n_1}(\mu_0,\eta_1)\chi_{n_2}(\kappa-\mu_0,\eta_2)
\end{equation}
If there are several minima of $B$ with the same value, one should take all
of them into account; expression (\ref{46}) should be replaced then by the
sum of analogous expressions associated with corresponding minima.

Let us make use of the explicit form of the functions $\chi_n,S$ and
simplify eq.(\ref{46}). Convert from quantities $\mu,\eta_1$ to parameters
$p_{\kappa_1},\tau_1$ and from quantities $\kappa-\mu,\eta_2$ to parameters
$p_{\kappa_2},\tau_2$ according to eq.(\ref{22}). The condition $
B^{'}(\mu_0)=0$  means that $p_{\kappa_1}=p_{\kappa_2}=p_{\kappa}$. Therefore,
in order to find the $\rho^{n_1n_2}_{k}$ asymptotics under conditions
(\ref{44*}) one must first find the parameters $p_{\kappa,\tau_1,\tau_2}$
according to the following relations,
\begin{equation}\label{47}
\kappa=(\lambda(\tau_1)+\lambda(\tau_2))e^{-p_{\kappa}},
\eta_1=Q(\tau_1)e^{-p_{\kappa/2}},
\eta_2=Q(\tau_2)e^{-p_{\kappa/2}}.
\end{equation}
The quantity $B(\mu_0)$ has then the form
$$
B_0=(s(\tau_1)+s(\tau_2)+(\lambda(\tau_1)+\lambda(\tau_2))p_{\kappa})
e^{-p_{\kappa}},
$$
while the second derivative of it is as follows,
\begin{equation}\label{47*}
B^{''}(\mu_0)=
\frac{\partial p_{\kappa_1}(\mu_0,\eta_1)}{\partial\mu_0} -
\frac{\partial p_{\kappa_2}(\kappa-\mu_0,\eta_2)}{\partial\mu_0} =
\frac{\partial \eta_1/\partial \tau_1}{J(\tau_1,p_{\kappa})}+
\frac{\partial \eta_2/\partial \tau_2}{J(\tau_2,p_{\kappa})},
\end{equation}
where $J$ is the Jacobian (\ref{32*}). Collecting all  obtained  factors,
one can find that the $\rho^{n_1n_2}_k$ asymptotics under conditions
(\ref{44*}) has the form of a product of a power function of $N$,
slowly varying pre-exponential factor and rapidly varying exponent:
\begin{equation}\label{48}
\rho^{n_1n_2}_{k}\sim N^{k+\frac{n_1+n_2-1}{2}}
\exp(-NB_0) \gamma,
\end{equation}
where
$$
\gamma=\frac{e^{-\frac{n_1+n_2-1}{2}p_{\kappa}}
e^{(n_1+1/2)\tau_1+(n_2+1/2)\tau_2}}{\sqrt{2\pi}\sqrt{
D}},
$$
$$
D=\dot{Q}(\tau_1)
[Q(\tau_2)\dot{\lambda}(\tau_2)/2-\lambda(\tau_2)\dot{Q}(\tau_2)]
+
\dot{Q}(\tau_2)
[Q(\tau_1)\dot{\lambda}(\tau_1)/2-\lambda(\tau_1)\dot{Q}(\tau_1)],
$$
the quantities $p_{\kappa},\tau_1,\tau_2$ are to be expressed from
 eq.(\ref{47}).

\subsection{High order asymptotics of some matrix elements}

This subsection
deals with the following example of application of the obtained
asymptotics formula (\ref{48}). Large order behaviour of perturbation
theory for matrix elements such as \begin{equation}\label{50}
<n_2|x^{m_1}(-d/dx)^{m_2}|n_1>=\sum_{k=0}^{\infty} g^k
<n_2|x^{m_1}(-d/dx)^{m_2}|n_1>_k,
\end{equation}
where $n_1$ and $n_2$ are numbers of levels, is studied. It occurs that
calculation of such asymptotics is based on the behaviour of eq.(\ref{48})
at coinciding arguments,
\begin{equation}\label{51}
\eta_1=\eta_2=\eta.
\end{equation}
Namely, the $k$-th order of perturbation theory for the quantity (\ref{50})
can be rewritten in a form,
\begin{equation}\label{51*}
\int dx dy \delta(x-y) x^{m_1}(-d/dx)^{m_2} \rho^{n_1n_2}_k(x,y).
\end{equation}
The $m_2$-th derivative of $\rho$ with respect to $x$ is approximately
equal to the product of $\rho$ by
$(\sqrt{N}\partial B_0/\partial \eta_1)^{m_2}$. Therefore, it is necessary
to know the functions $B_0,\gamma$ only at the surface (\ref{51}).

Let us now find the parameters $\tau_1,\tau_2,p_{\kappa}$ when eq.(\ref{51})
is satisfied. Notice first that one of possible choices of $\tau_1,\tau_2$
is as follows,$\tau_1=\tau_2=\tau$, and, therefore,
\begin{equation}\label{52}
\eta/\sqrt{\kappa}=Q(\tau)/\sqrt{2\lambda(\tau)}.
\end{equation}
This choice corresponds to the following selection of the parameter $\mu$
in eq.(\ref{45*}),$\mu=\kappa/2$. It is always the extremum of the
function $B(\mu)$. Eq.(\ref{47*}) implies that this extremum is minimum when
$\dot{Q}>0$ (fig.6a) and maximum when $\dot{Q}<0$ (fig.6b). Therefore, if the
parameter $\tau$ obtained from eq.(\ref{52}) corresponds to negative values of
the velocity then it is necessary to find other extremum of $B(\mu)$ that is
minimum.

Another possibility is as follows.
Since the trajectory $Q(\tau)$ passes through each point twice, one can
consider the parameters $\tau_1,\tau_2$ such that $Q(\tau_1)=Q(\tau_2)$ but
$\dot{Q}(\tau_1)>0,\dot{Q}(\tau_2)<0$ (see fig.6c),so that
\begin{equation}\label{53}
\eta/\sqrt{\kappa}=Q(\tau_1)/\sqrt{s(+\infty)}.
\end{equation}
This selection is possible when
$\eta/\sqrt{\kappa}<Q_+/\sqrt{s(+\infty)}$,
contrary to the previous case being possible as
$\eta/\sqrt{\kappa}>Q_+/\sqrt{s(+\infty)}$,
and corresponds to minimum of $B$, because $J(\tau_2)-J(\tau_1)=
e^{-3p_{\kappa}/2}\dot{Q}(\tau_1)s(+\infty)>0$ and, thereby,
$B^{''}(\mu_0)>0$. Take also into account that we can exchange the variables
$\mu$ and $\kappa-\mu$ and obtain another minimum of $B$.

Making use of the cited observations, let us find the constants $B,\gamma$
in the
\newline
$\rho^{n_1n_2}_{N\kappa}(\sqrt{N}\eta,\sqrt{N}\eta)$ asymptotics:

(a) $\eta/\sqrt{\kappa}>Q_+/\sqrt{s(+\infty)}:$
$$
B=\kappa(\frac{s(\tau)}{\lambda(\tau)}+\ln\frac{2\lambda(\tau)}{\kappa}),
$$
$$
\gamma=\frac{1}{\sqrt{2\pi}}
\left(\frac{\kappa}{2\lambda}\right)^{(n_1+n_2-1)/2}
\frac{e^{(n_1+n_2+1)\tau}}{ \sqrt{2\dot{Q}(\tau)[Q(\tau)\dot{\lambda}(\tau)/2-
\lambda(\tau)\dot{Q}(\tau)] }}.
$$
Notice that in this case $\dot{B}=-\dot{Q}J/\lambda^2<0$, so that as $\tau$
decreases, $B$ increases. Therefore, the main contribution to the integral
(\ref{51*}) is given by a region near the critical value of
$\eta/\sqrt{\kappa}$, $Q_+/\sqrt{s(+\infty)}$.

(b) $\eta/\sqrt{\kappa}<Q_+/\sqrt{s(+\infty)}:$
$$
B=\kappa(1+\ln\frac{s(+\infty)}{\kappa}),
$$
\begin{equation}\label{54}
\gamma=\frac{(\kappa/s(+\infty))^{(n_1+n_2-1)/2} }
{2\pi s(+\infty)\dot{Q}(\tau_1)}
[e^{(n_1+1/2)\tau_1+(n_2+1/2)\tau_2}+
e^{(n_1+1/2)\tau_2+(n_2+1/2)\tau_1}],
\end{equation}
 the existence of two minima
of the function $B$ is taken into account.  One can notice that the
region (a) gives an exponentially small contribution to
eq.(\ref{51*}).

Find now large order asymptotics of the matrix element (\ref{50}). Its
evaluation is based on the following observations:

(i) one can change the parameter $x$ by a parameter $\tau_1$, according
to eq.(\ref{41}), the measure $dx$ transforms then to
$\sqrt{N}\dot{Q}(\tau_1)e^{-p_{\kappa}/2}d\tau_1$, while the
derivation operator $-d/dx$ should be substituted by the operator of
multiplication by $\sqrt{N}P(\tau_1)e^{-p_{\kappa}/2}$ at the first
term in eq.(\ref{54}) and by $\sqrt{N}P(\tau_2)e^{-p_{\kappa}/2}$ at
 the second;

(ii) the following relations are satisfied: $Ne^{-p_{\kappa}}=k/s(+\infty)$,
$\tau_1+\tau_2=\ln c$, where the parameter $c$ has the form (\ref{35c});

(iii) one can divide the obtained integral over $\tau_1$, where
$\tau_1<\tau_0,Q(\tau_0)=Q_+$ into two parts according to eq.(\ref{54}),
denote by $\tau$ the quantity $\tau_1$ in the first integral and $\tau_2$ in
the second one, the rest parameter is then equal to $\ln c-\tau$; the
integrands become coinciding then, while the region of integration is
$\tau<\tau_0$ in the first integral and $\tau>\tau_0$ in the second, so
their sum is an integral over $-\infty<\tau<+\infty$;

(iv) one can employ the asymptotic formula for $k!$;

(v) notice that at odd $m_1+m_2+n_1+n_2$ the quantity (\ref{50}) is obviously
equal to zero, while at even $m_1+m_2+n_1+n_2$ the integrals over regions
$x<0$ and $x>0$ coincides, so one should increase the value of the integral
over $x>0$ in 2 times and obtain the following asymptotic formula:
$$
<n_2|x^{m_1}(-d/dx)^{m_2}|n_1>_k \sim \frac{\Gamma(k)}{\pi
(s(+\infty))^k}\left(\frac{k}{s(+\infty)}\right)^{\frac{n_1+n_2+m_1+m_2}{2}}
$$
\begin{equation}\label{55}
\times\int_{-\infty}^{+\infty} d\tau e^{(n_2+1/2)\ln
c}Q^{m_1}(\tau)P^{m_2}(\tau)e^{(n_1-n_2)\tau},
 \end{equation}
Let us discuss the obtained result (\ref{55}). Notice first that if we
substitute $\tau$ by $\ln c -\tau$, this expression will be multiplied by
$(-1)^{m_2}$, while the numbers $n_1$ and $n_2$ will exchange. This fact is
in agreement with the definition (\ref{50}).

Second, notice that when $m_1$ and $m_2$ increase, the growth of the
coefficients is faster. The $(k-1)$-th order for the matrix element of $x^m$
and the $k$-th order for such quantity for $x^{m-2}$ have the same order of
growth. Therefore, the $k$-th order behaviour for all powers of
$gx^2,g(d/dx)^2$ is of the same order.

Large order asymptotics for quantities such as ''Green functions''
\begin{equation}\label{56}
<n_2|x(\tau_1,...,\tau_m|n_1>_k,
\end{equation}
where $x(\tau)$ is an operator of the form
$x(\tau)=e^{{\cal H}\tau}xe^{-{\cal H}\tau}$ can be also studied. Namely,
denote the solution to eq.(\ref{13*}) satisfying initial conditions
$P(0)=P_0,Q(0)=Q_0$ by $P(\tau,Q_0,P_0),Q(\tau,Q_0,P_0).$
The operator $x(\tau)$ is then approximately equal to
$$
x(\tau)\sim g^{-1/2} Q(\tau,\sqrt{g}x,-\sqrt{g}d/dx).
$$
One can expand this function into a series in operators $x$ and $-d/dx$
and apply eq.(\ref{55}). It occurs then that large order asymptotics for
eq.(\ref{56}) can be obtained from eq.(\ref{55}) with the help of substituting
$m_1$ by $m$,$m_2$ by 0, $Q^m(\tau)$ by a product
$Q(\tau+\tau_1)...Q(\tau+\tau_m).$This result is in agreement with the
Lipatov method \cite{L}
of finding asymptotics for such quantities as (\ref{56})
for the ground state, namely, eq.(\ref{56}) can be presented as an integral
over trajectories
$$
\int Dx \oint \frac{dg}{2\pi i g^{k+1}} x(\tau_1)...x(\tau_m)
e^{-s(\sqrt{g}x)/g};
$$
after substitution $x\sqrt{g}=Q,g=\lambda/k$ this integral becomes of the
saddle-point type, saddle points being $Q=Q(\tau+\alpha),\lambda=s(+\infty)$
are to be taken into account, so the asymptotics of eq.(\ref{56}) is expressed
through the integral $\int d\alpha Q(\tau_1+\alpha)...Q(\tau_n+\alpha).$

\section{Conclusions}

In this paper the explicit mechanism showing the connection between large
order recursive relations for stationary perturbation theory and classical
euclidean equations of motion is found and high order asymptotics of the
solution to the recursive relations is constructed. The results obtained by
the path integral approach are indeed reproduced in the developed method.
Moreover, when one finds high order asymptotics of matrix element
(\ref{50}), one can also find what region of integration in eq.(\ref{51*})
and what terms in the sum (\ref{44}) give the main contribution to the
quantity (\ref{50}). Making use of the obtained large order asymptotics
for the wave function and density matrix, one can also find the main values
of quantities like $e^{-kf(x/\sqrt{k})}$ at the $k$-th order of perturbation
theory as $k\rightarrow\infty$. All calculations are valid both for the
ground and for the excited states.

An interesting feature of the eigenfunction large order perturbation theory
as $x/\sqrt{k}=const$ is the divergence of the pre-exponential factor
at small $\xi$. This difficulty is analogous with the divergence of the
pre-exponential factor in the semiclassical expansion large order asymptotics
discussed in  \cite{S1}. The resolve of the difficulty is
the following. At small $\xi$ associated with the values of parameter $x$
of order $O(1)$ the asymptotics is written in another form (\ref{38}).
 It is expressed through the growing at infinity solution to the harmonic
 oscilator equation with non-trivial right-hand side (\ref{41}).
  The connection
 between these asymptotics allows us to find the eigenvalue large order
 behaviour coinciding with previous papers \cite{ZJ}.

The author is indebted to V.A.Rubakov for helpful discussions.
The work is supported in part by ISF, grant \# MKT000.

\newpage

{\bf Figure captions}

{\bf Fig.1.}

(a): The potential $V$ versus $Q$;

(b): classical euclidean solution in phase space; the area of the
section-lined region is equal to the quantity $\lambda$, eq.(\ref{22*});

(c): the correspondence between the parameter $\xi$, eq.(\ref{23}), and
the coordinate $Q$.

{\bf Fig.2.}

A set of classical trajectories in $\kappa-\eta$ space.

{\bf Fig.3.}

The functions $A_k(\xi)$ defined by eq.(\ref{A}) at various numbers of the
order of perturbation theory,$k=10,20,30$, are shown by dashed lines;
the function $A(\xi)$, eq.(\ref{28}), is shown by solid line.

{\bf Fig.4.}

The ratio $M_k(\xi)$, eq.(\ref{MK}), at $k=10,20,30,40$ and the function
$M(\xi)$ specified by eq.(\ref{M}).

{\bf Fig.5.}

The functions $X_{0,k}(x)$ defined by eq.(\ref{X}) at $k=10,20,30,40$
(solid lines) and the solution to eq.(\ref{41}) (dashed line).

{\bf Fig.6.}

Classical trajectories in $\kappa-\eta$ space which correspond to various
choices of the parameter $\mu$, an extremum point of the function (\ref{45*}):

(a) $\mu=\kappa/2,\dot{Q}(\tau)>0$;

(b) $\mu=\kappa/2,\dot{Q}(\tau)<0$;

(c) $\mu\ne\kappa/2$.


\newpage
\begin{thebibliography}{99}
\bibitem{D} F.J.Dyson  {\em Phys. Rev.} {\bf 85} (1952) 631.
\bibitem{L}  L.N.Lipatov. {\em Sov. Phys. JETP} {\bf 45} (1977) 216.
\bibitem{BW2} C.M.Bender, T.T.Wu {\em Phys.Rev.} {\bf D7} (1973) 1620.
\bibitem{BBW} T.Banks, C.M.Bender, T.T.Wu {\em Phys.Rev.}
 {\bf D8} (1973) 3346.
\bibitem{BLGZJ} E.Br\'{e}zin, J.C.Le Guoillou, J.Zinn-Justin {\em Phys.Rev.}
                  {\bf D15} (1977) 1558.
\bibitem{BPZJ} E.Br\'{e}zin, G.Parisi, J.Zinn-Justin {\em Phys.Rev.} {\bf D16}
 (1977) 408.
\bibitem{ZJ} J.Zinn-Justin {\em Phys.Rep.}{\bf 70} (1981) 109.
\bibitem{RS} V.A.Rubakov, O.Yu.Shvedov. {\em Nucl.Phys.} {\bf B434} (1995) 245.
\bibitem{S2} O.Yu.Shvedov, quant-ph/9504014.
\bibitem{BW1} C.M.Bender, T.T.Wu {\em Phys.Rev.} {\bf 184} (1969) 1231.
\bibitem{M} V.P.Maslov. Asymptotic methods and perturbation theory.
      Moscow,Nauka,1988.
\bibitem{S1} O.Yu.Shvedov. quant-ph/9504012, submitted to Yadernaya Fizika.

\end{thebibliography}
\end{document}